\documentstyle[multicol,aps,prl,epsf]{revtex}
\begin{document}
\title{Global Persistence Exponent for Critical Dynamics}
\author{S. N. Majumdar$^1$, A. J. Bray$^2$, S. J. Cornell$^2$, and C. Sire$^3$}
\address{
$^1$ Physics Department, Yale University, New Haven, CT 06520-8120 \\
$^2$ Department of Theoretical Physics, The University, Manchester M13 9PL, 
UK \\
$^3$ Laboratoire de Physique Quantique (UMR C5626 du CNRS), Universit\'e 
Paul Sabatier, 31062 Cedex, France
}
\date{June 17, 1996}
\maketitle

\begin{abstract}
A `persistence exponent' $\theta$ is defined for nonequilibrium critical 
phenomena. It describes the probability, $p(t) \sim t^{-\theta}$, that the 
global order parameter has not changed sign in the time interval $t$ following 
a quench to the critical point from a disordered state. This exponent is 
calculated in mean-field theory, in the $n=\infty$ limit of the $O(n)$ model, 
to first order in $\epsilon = 4-d$, and for the 1-d Ising model. Numerical 
results are obtained for the 2-d Ising model. We argue that $\theta$ is 
a new independent exponent. 

\end{abstract}

%\narrowtext

%\pacs{}
\begin{multicols}{2}
For many years it was believed that critical phenomena were characterized 
by a set of three critical exponents, comprising two independent static 
exponents (other static exponents being related to these by scaling laws) 
and the dynamical exponent $z$. Then, quite recently, it was discovered 
that there is another dynamical exponent, the `non-equilibrium' (or 
`short-time') exponent $\lambda$, needed to describe two-time correlations 
in a system relaxing to the critical state from a disordered initial 
condition \cite{Janssen,Huse}. It is natural to ask `Are there any more 
independent critical exponents?'. In this Letter we propose such an 
exponent -- the `persistence exponent' $\theta$ associated with the 
probability, $p(t) \sim t^{-\theta}$, that the global order parameter 
has not changed sign in time $t$ following a quench to the critical point. 
We calculate $\theta$ in mean-field theory, in the $n=\infty$ limit of the 
$O(n)$ model, to first order in $\epsilon=4-d$ ($d=$ dimension of space) 
and for the $d=1$ Ising model. In fact, it turns out that all these results 
satisfy the scaling law $\theta z = \lambda - d + 1 - \eta/2$, which can 
be derived on the assumption that the dynamics of the global order parameter 
is a Markov process. We shall argue, however, that this process is in 
general non-Markovian, so that $\theta$ is in general a new, non-trivial 
critical exponent.  

The persistence exponent $\theta$ was first introduced in the context  
of the non-equilibrium coarsening dynamics of systems at zero temperature 
\cite{Derrida94,Bray94}. In that context it describes the power-law 
decay, $p(t) \sim t^{-\theta}$, of the probability that the local order 
parameter $\phi({\bf x})$ has not changed sign during the time interval 
$t$ after the quench to $T=0$. Equivalently, it gives the fraction of 
space in which the order parameter has not changed sign up to time $t$. 
More generally, one can consider the probability $p_0(t_1,t_2)$ of no 
sign changes between $t_1$ and $t_2$. Scaling considerations 
suggest $p_0(t_1,t_2) = f(t_1/t_2) \sim (t_1/t_2)^\theta$ for $t_2 \gg t_1$. 
 
Exact solutions for one-dimensional systems \cite{Bray94,DHP} indicate  
that, in general, $\theta$ is a new non-trivial exponent for coarsening 
dynamics. Recently, we have shown that even the diffusion equation 
exhibits a nontrivial persistence exponent, and have developed a  
rather accurate approximate theory for this case \cite{MSBC}. The diffusion 
equation is itself a model of ordering dynamics, via the approximate 
theory of Ohta, Jasnow and Kawasaki (OJK) \cite{OJK}, and also describes, 
in its essential features, the ordering kinetics of the nonconserved $O(n)$ 
model in the large-$n$ limit \cite{Review}: The exponents $\theta$ for these 
systems (OJK and large-$n$) are just those of the diffusion equation.   

In this Letter we introduce and calculate the analogous exponent 
$\theta$ for non-equilibrium {\em critical} dynamics. 
In this case however, one needs to consider the {\em global}, rather than 
the {\em local} order parameter. This is because individual degrees of 
freedom (`spins', say) are rapidly flipping so that the probability of not 
flipping in an interval $t$ has an exponential tail. We shall see, 
however, that the probability for the {\em global} order parameter not to 
have flipped indeed decays as a power law. One simplifying property of 
the global order parameter is that, in the thermodynamic limit, it 
remains Gaussian at all finite times. This follows from the central 
limit theorem, on noting that the order-parameter field 
$\phi({\bf x},t)$ has a finite correlation length, $L(t) \sim t^{1/z}$. 
If the system has a volume $V \gg L(t)^d$, the appropriate 
Gaussian variable is the $k=0$ Fourier component, 
$\tilde{\phi}_0(t)=[\int d^dx\,\phi({\bf x},t)]/\sqrt{V}$. From standard 
scaling, $\langle \tilde{\phi}_0^2(t) \rangle \sim L(t)^{2-\eta}$. This 
follows from the $k \to 0$ limit of the scaling form \cite{Janssen}
$\langle\tilde{\phi}_{\bf k}(t)\tilde{\phi}_{-\bf k}(t)\rangle 
 = k^{-(2-\eta)}G[kL(t)]$. 

Our explicit results are derived from the Langevin equation for the vector 
order parameter $\vec{\phi} = (\phi_1,\ldots,\phi_n)$:
\begin{equation}
\partial_t \phi_i = \nabla^2 \phi_i - r\phi_i - (u/n)\phi_i\sum_j\phi_j^2\ 
+ \xi_i\ ,
\label{Langevin}
\end{equation}
where $\vec{\xi}(\bf{x},t)$ is a Gaussian white noise with mean zero and 
correlator $\langle \xi_i({\bf x},t)\xi_j({\bf x}',t')\rangle = 
2 \delta_{ij}\,\delta^d({\bf x}-{\bf x}')\,\delta(t-t')$. [For a vector 
order parameter, we are defining $p(t)$ as the probability that a given 
{\em component} of the global order parameter,  
$\vec{\Phi}(t) =\int d^dx\,\vec{\phi}({\bf x},t)$, has not changed sign]. 

In mean-field theory, valid for $d \ge 4$, we set $r=0=u$. Then the 
${\bf k}=0$ Fourier component $\tilde\phi_i(0,t) = \Phi_i(t)/\sqrt{V}$ 
(where $V$ is the volume) obeys the simple equation (suppressing the 
index $i$ and the arguments) 
\begin{equation}
\partial_t \tilde{\phi} = \tilde{\xi}\ ,
\label{mean-field}
\end{equation}
indicating that $\tilde\phi$ executes a simple random walk. The 
non-flipping probability $p(t)$ is therefore just the probability that the 
random walker has not crossed the origin up to time $t$. It is given by 
\cite{rw} $p(t) \sim |\tilde{\phi}_0|/\sqrt{t}$ for large $t$, where 
$\tilde{\phi}_0$ is the initial value of $\tilde{\phi}$. Finally one has 
to average over $|\tilde{\phi}_0|$. For a disordered initial condition, 
$\Phi(0) \sim \sqrt{V}$ by the central limit theorem, so 
$\tilde{\phi}_0 = \Phi(0)/\sqrt{V}$ is $O(1)$, 
the desired average is also $O(1)$, and $p(t) \sim 1/\sqrt{t}$. 
We conclude that $\theta=1/2$ in mean-field theory. 

Next we consider the large-$n$ limit. Equation (\ref{Langevin}) 
then simplifies to the self-consistent linear equation 
\begin{equation}
\partial_t \phi = \nabla^2 \phi - (r + u\langle \phi^2\rangle)\phi + \xi ,
\label{large-n}
\end{equation}
for each component. Defining $a(t) = - r - u\langle \phi^2 \rangle$ and 
$b(t) = \int_0^t a(t')\,dt'$, (\ref{large-n}) has the Fourier-space 
solution
\begin{eqnarray}
\tilde{\phi}({\bf k},t) & = & \tilde{\phi}(0,t)\exp[b(t)-k^2t] \nonumber\\
&+& \int_0^t dt'\,\tilde{\xi}({\bf k},t')\exp[b(t)-b(t')-k^2(t-t')]\ .
\end{eqnarray} 
It is easy to show that the second term, involving the noise, dominates 
the first at large $t$ \cite{Janssen}. Retaining only the second, 
computing $\langle \phi^2 \rangle$, and defining $g = \exp(-2b)$ leads 
to the equation
\begin{equation}
\partial_tg = 2rg + 4u \int_0^t dt'\,g(t')\sum_{\bf k} 
\exp[-2k^2(t-t')]\ , 
\end{equation}
which can be solved by Laplace transformation. Setting $r$ equal to its 
critical value, $r_c = -u\sum_{\bf k} k^{-2}$ gives 
\begin{eqnarray}
\bar{g}(s) & = & [s + 4u \{\bar{J}(0)- \bar{J}(s)\}]^{-1}\ , \\
\bar{J}(s)    & = & \sum_{\bf k} (s + 2k^2)^{-1}\ , 
\end{eqnarray}
from which one deduces $\bar{g} \sim s^{(2-d)/2}$ for $s \to 0$, 
for $2<d<4$. Inverting the Laplace transform gives (with $\epsilon = 4-d$)
$g(t) \sim t^{-\epsilon/2}$ for $t \to \infty$, whence 
$b \sim (\epsilon/4)\ln t$, and $a(t) \sim \epsilon/4t$. 

The large-$n$ equation of motion (\ref{large-n}) therefore reduces to 
\begin{equation}
\partial_t \tilde{\phi} = (\epsilon/4t)\tilde{\phi} + \tilde{\xi}
\label{simple}
\end{equation}
for the ${\bf k}=0$ Fourier component of $\phi$. The final step is to 
eliminate the first term on the right by setting 
$\tilde{\phi} = t^{\epsilon/4}\psi$, to give 
$\partial_t \psi = t^{-\epsilon/4}\tilde{\xi}(t)$. Introducing the new time 
variable $\tau = t^x$, this equation reduces to the random walk equation 
$\partial_\tau \psi = \eta(\tau)$, with $\eta$ a Gaussian white noise, if 
one chooses $x = (2 - \epsilon)/2$. The final result is therefore 
$p(t) \sim \tau^{-1/2} = t^{(2-d)/4}$, giving
\begin{equation}
\theta = (d-2)/4, \ \ \ \ \ 2<d<4 \ \ (n=\infty)\ .
\label{largen-result}
\end{equation}  
For $d>4$, $\theta$ sticks at the mean-field value of $1/2$. 

Finally, we calculate $\theta$ to first order in $\epsilon=4-d$. This is 
most simply accomplished using the method of Wilson \cite{Wilson}. 
To order $\epsilon$ the calculation can be carried out in $d=4$, by expanding 
$p(t)$ to first order in $u$, setting $u$ equal to its Renormalization-Group 
(RG) fixed-point value, and exponentiating logarithms. 

The perturbative calculation of $p(t)$ is in principle quite a difficult 
task. A systematic technique for performing the perturbation expansion  
was recently developed by two of us \cite{MS} in the general 
context of first-passage-time problems for non-Markov Gaussian processes. 
It amounts to expanding around the random walk (2) within a path-integral 
formulation of the problem. Since the global order-parameter $\Phi(t)$ 
remains Gaussian at all times (in the thermodynamic limit), this method is 
applicable. In the present work, however, we restrict ourselves to first 
order in $\epsilon$, for which the result can be obtained by elementary 
methods. The reason is that the dynamics of $\Phi(t)$ remain Markov to 
this order, as we shall see. 

First we replace $u/n$ in (\ref{Langevin}) by $u$, to conform to 
the conventional notation for $\epsilon$-expansions. To first order in $u$, 
one can as usual replace the nonlinear term $u\phi_i\sum_j\phi_j^2$ in 
(\ref{Langevin}) by the linearized form 
$(n+2)u\langle \phi_j^2 \rangle \phi_i$. The required expression for 
$\langle \phi_j^2 \rangle$ can be evaluated at $u=0$ and $r=0$, since $r_c$ 
is also $O(u)$. For this part of the calculation, therefore, we can use 
the mean-field result, 
\begin{equation}
\tilde{\phi}_{\bf k}(t) = \tilde{\phi}_{\bf k}(0)\exp(-k^2t)  
+ \int_0^t dt'\, \tilde{\xi}_{\bf k}(t')\,\exp[-k^2(t-t')],
\end{equation}  
to give 
\begin{equation}
\langle \phi_j^2 \rangle = \Delta \sum_{\bf k}\exp(-2k^2t) 
+ \sum_{\bf k} \frac{1}{k^2} \left(1 - \exp(-2k^2t)\right),
\end{equation}
where we have used 
$\langle \tilde{\phi}_{\bf k}(0)\tilde{\phi}_{-\bf k}(0)\rangle = \Delta$, 
appropriate to short-range spatial correlations in the initial state. 

To order $u$ the critical point is $r_c = -(n+2)u\sum_{\bf k} k^{-2}$. The 
effective Langevin equation for the $k=0$ mode, correct to $O(u)$, is 
therefore (suppressing the index $i$ and the momentum subscript) 
\begin{equation}
\partial_t \tilde{\phi} = (n+2)u\sum_{\bf k}
\left(\frac{1}{k^2}-\Delta\right)\exp(-2k^2t)\,\tilde{\phi} + \tilde{\xi}.
\end{equation}
The ${\bf k}$-integrals are now performed in $d=4$. It is clear that the 
term involving the initial-condition correlator $\Delta$ is smaller 
(by one power of $t$) than that coming from the thermal noise, and may 
therefore be dropped, giving 
\begin{equation}
\partial_t \tilde{\phi} = (n+2)\frac{uK_4}{4t}\,\tilde{\phi} + \tilde{\xi},
\label{epsilon}
\end{equation}
where $K_4 = 1/8\pi^2$ is the usual geometrical factor. Setting $u$ equal 
to its RG fixed-point value \cite{Wilson} $u^* = \epsilon/[(n+8)K_4]$ gives 
an equation identical to the large-$n$ equation (\ref{simple}), but with the 
replacement $\epsilon \to [(n+2)/(n+8)]\epsilon$. Making the same replacement 
in (\ref{largen-result}), we deduce immediately that the exponent $\theta$ 
is given by 
\begin{equation}
\theta = \frac{1}{2} - \frac{1}{4}\left(\frac{n+2}{n+8}\right) \epsilon 
          + O(\epsilon^2), 
\label{epsilon1}
\end{equation}
which agrees with (\ref{largen-result}) for $n \to \infty$. 
For the Ising universality class ($n=1$), (\ref{epsilon1}) becomes 
$\theta = 1/2 - \epsilon/12 + O(\epsilon^2)$. 

The final soluble limit we consider is the $d=1$ Ising model with Glauber 
dynamics. For this model, the critical point is at $T=0$, so the `critical' 
and `strong coupling' fixed points coincide. The persistence probability 
$p(t)$ for a single spin has been considered earlier \cite{Bray94}. Very 
recently, it has been shown to decay as $t^{-3/8}$, with non-trivial results 
for general $q$-state Potts models \cite{DHP}. The calculation of $p(t)$ 
for the global magnetization $M(t)$ is much simpler. At $T=0$ the dynamics 
is equivalent to a set of annihilating random walkers (the domain walls). 
At each time step, each random walker moves independently one step to 
the left or right \cite{Note2}. Let the number of spins be $N$. Then the 
number of surviving walkers at time $t$ is of order $Nt^{-1/2}$ 
\cite{Bray90,Amar}. The change in $M(t)$ in one time step is therefore of 
order $\sqrt{N} t^{-1/4}$, since the contributions from the walkers add 
incoherently. The ${k=0}$ Fourier component $\tilde{\phi} = M/\sqrt{N}$ 
therefore satisfies the Langevin equation (up to constants) \cite{MH}
\begin{equation}
\partial_t \tilde{\phi} = t^{-1/4} \xi(t),
\label{1d}
\end{equation} 
where $\xi(t)$ is a Gaussian white noise, $\langle \xi(t)\xi(t') \rangle 
= C\delta(t-t')$, and $C$ is a constant.

This can be reduced to the standard random-walk dynamics through the change 
of variable $t=\tau^2$. Equation (\ref{1d}) then reads $d\tilde{\phi}/d\tau 
= 2\tau^{1/2}\xi(\tau^2) \equiv \eta(\tau)$, where $\eta(\tau)$ has correlator 
$\langle \eta(\tau) \eta(\tau') \rangle = 4C\tau\delta(\tau^2-\tau'^2) 
= 2C\delta(\tau-\tau')$, i.e.\ $\eta(t)$ is a Gaussian white noise. We conclude 
that $p(t) \propto \tau^{-1/2} = t^{-1/4}$, i.e.\ $\theta = 1/4$ for this model.
It is remarkable, but certainly coincidental, that the $O(\epsilon)$ result 
gives this result exactly, on setting $\epsilon=3$. 

At this point we note a simplifying feature of all the results
presented so far, namely the underlying dynamics is a Gaussian Markov process 
in every case. This should be apparent from equations (\ref{mean-field}), 
(\ref{simple}), (\ref{epsilon}), and (\ref{1d}). For such cases one can 
derive (see below) a scaling law relating $\theta$ to other exponents, namely   
\begin{equation}
\theta z = \lambda - d + 1 - \eta/2\ ,
\label{scaling}
\end{equation}
where $\lambda$ describes the asymptotics of the two-time correlation 
function of the global order parameter at $T_c$: 
$\langle \tilde{\phi}(t_1) \tilde{\phi}(t_2)\rangle = 
t_1^{(2-\eta)/z} F(t_2/t_1)$, with $F(x) \sim x^{(d-\lambda)/z}$ for 
$x \to \infty$. Using the known results $\eta=0$, $z=2$, 
$\lambda = (3d-4)/2$  for $n=\infty$ \cite{Janssen}, $\eta=0$, $z=2$, 
$\lambda = d - [(n+2)/(n+8)]\epsilon/2$ to $O(\epsilon)$ \cite{Janssen}, 
and $\eta=1$, $z=2$, $\lambda=1$ for the $d=1$ Ising model \cite{Bray90}, 
one can check that all of the results derived above satisfy this scaling law. 

Does this scaling law hold generally? We do not think so: we believe that  
$\Phi(t)$ is {\em not} a Markov process in general (though it is Gaussian), 
for the following reason. Consider the autocorrelation function for the 
$k=0$ mode, $\langle \tilde{\phi}(t_1)\tilde{\phi}(t_2) \rangle$. We have 
seen that it has the scaling form $t_1^{(2-\eta)/z}F(t_2/t_1)$, with 
$F(x) \sim x^{(d-\lambda)/z}$ for $x \to \infty$. Now construct the 
normalized autocorrelation function $a(t_1,t_2) = 
\langle \tilde{\phi}(t_1)\tilde{\phi}(t_2) \rangle/
\langle \tilde{\phi}(t_1)^2\rangle^{1/2} 
\langle \tilde{\phi}(t_2)^2 \rangle^{1/2}$. This has the scaling form 
$a(t_1,t_2) = f(t_1/t_2)$, with $f(x) \sim x^{(\lambda-d+1-\eta/2)/z}$ for 
$x \to \infty$. If we introduce the new time variable $T=\ln t$, this becomes 
$A(T_1,T_2) = g(T_1-T_2)$, i.e.\ the process is  a Gaussian stationary  
process in this time variable. Also the function $g(T)$ has the asymptotic 
form $g(T) \sim \exp(-\bar{\lambda} |T|)$, with $\bar{\lambda} = 
(\lambda-d+1-\eta/2)/z$. 

Now if the process is Markovian, $g(T)$ necessarily has this exponential 
form for {\em all} $T$, not just asymptotically large $T$ \cite{GM}. 
Futhermore, the first-passage exponent $\theta$ is then equal to 
$\bar{\lambda}$ \cite{rw,GM,GM1}, which is the origin of the scaling law 
(\ref{scaling}) for Markov processes. Note that, in the original time 
variables, requiring $g(T)$ to be a simple exponential is equivalent to 
requiring that the scaling function $f(t_1/t_2)$ be a simple power of 
$t_1/t_2$ for {\em all} $t_2>t_1$, not just $t_2 \gg t_1$.  

So the question of whether $\Phi(t)$ is a Markov process comes down to the 
question of whether the scaling function $f(t_1/t_2)$ [of the normalized 
autocorrelation function of $\tilde{\phi}(t)$] is a simple power law 
for all $t_2 \ge t_1$.  The known results for $n=\infty$, $O(\epsilon)$, and 
the $d=1$ Ising model satisfy this requirement. For the last of these, the
reduction to a random walk makes this transparent. In the other two cases, 
it is consequence of the simplicity of the one-loop nature of the 
calculations, which give simple powers. To $O(\epsilon^2)$, however, the 
structure of the `watermelon' two-loop graph leads to a nontrivial 
dependence on $t_1/t_2$, which does not reduce to a simple power 
\cite{BCMS}. It follows that the putative scaling law (\ref{scaling}) 
will fail at $O(\epsilon^2)$: The dynamics of the global order parameter 
are non-Markovian in general, and the exponent $\theta$ is an independent 
critical exponent. Similar conclusions follow consideration of the next term
in the $1/n$ expansion. 

The exponent $\theta$ was measured numerically for 2-d Ising systems, using a 
finite-size scaling technique for square lattices of linear size $L =$ 24, 32, 
48, 64, 96 and 128, with periodic boundary conditions. Starting from a random 
initial condition, the systems were evolved using heat-bath Monte-Carlo 
dynamics at the bulk critical coupling $K_c = [\ln(1+\sqrt{2})]/2$. 
Each system was evolved until the global magnetization first changed sign. 
The fraction $p(t)$ of surviving systems was then computed at each time $t$, 
over a number of runs varying from $200\,000$ for $L=24$ to $91\,008$ 
for $L=128$. Finite-size scaling suggests the scaling form 
$p(t) = t^{-\theta} f(t/L^z) = L^{-\theta z} \bar{f}(t/L^z)$, where $z$ 
is the dynamic exponent. We therefore plot $L^{\theta z}p(t)$ against 
$t/L^z$, and vary $\theta$ for the best data collapse. The dynamic exponent 
was taken to be $z=2.15$ \cite{z}. Data for $t<20$ were discarded. 
The best collapse was obtained for $\theta z = 0.50 \pm 0.02$. Scaling plots 
for $\theta z=0.48$, 0.50 and 0.52 are shown in Figure 1. We have tried values 
of $z$ other than 2.15, but find that $z=2.1$ and 2.2 give significantly worse 
data collapse.  

A finite-size scaling analysis is essential here, as the data show significant 
curvature in the `early' time regime, even for the largest systems studied 
($128^2$). In the scaling form $L^{\theta z} p(t) = F(t/L^z)$, the scaling 
function $F(x)$ must vary as $x^{-\theta}$ for small $x$, but the `small-$x$' 
regime in the data is not extensive  enough to extract the exponent from this 
part of the plot alone. In the large-$x$ regime, one expects $F(x) \sim 
\exp(- const\,x)$, since $L^z$ is the characteristic relaxation time 
of the system. This behavior is confirmed in studies of smaller systems, 
where longer runs are feasible. The final part of the scaling plots 
in Figure 1 show the entry into this exponential regime.  

\begin{figure}[h]
\narrowtext
\epsfxsize=\hsize
\epsfbox{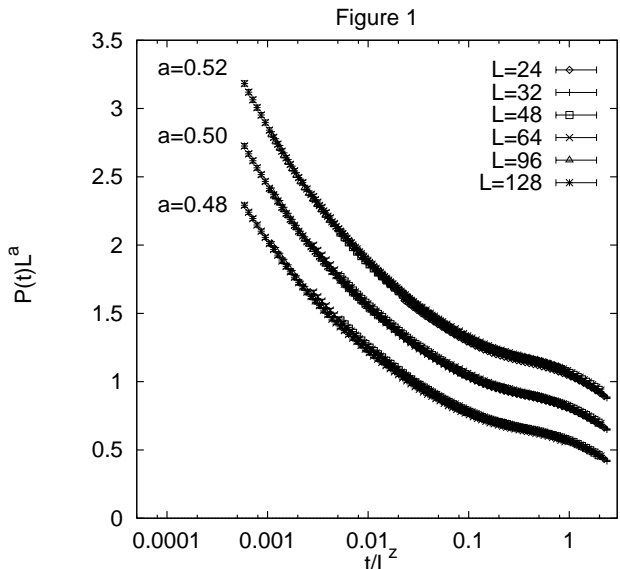} 
\caption{Finite-size scaling plots for the `persistence'  $p(t)$ 
(the fraction of systems whose total magnetization has not changed sign) for 
the $d=2$ Ising model at $T_c$, with $z=2.15$ and $a\equiv \theta z = $ 
0.48, 0.50 and 0.52. For clarity, the 0.50 and 0.52 data have been moved up 
by 0.2 and 0.4 respectively. }
\end{figure}

It is interesting to compare the numerical result for $\theta z$ with the 
prediction of the `scaling law' (\ref{scaling}).  Using 
$\lambda \approx 1.59 \pm 0.05$ \cite{Huse}, and the exact result $\eta=1/4$, 
(\ref{scaling}) gives $\theta z = 0.47 \pm 0.05$, consistent with the 
measured value $0.50 \pm 0.02$. This suggests that non-Markovian violations 
of the relation (\ref{scaling}) may be quite small in practice. 

In summary we have identified a new exponent $\theta$ for critical dynamics. 
It is the analog of the `persistence' exponents discussed in a number of other 
contexts recently, and characterizes the time-dependence of the probability 
that the global order parameter has not changed sign up to time $t$ after a 
quench to the critical point from the disordered phase. We have argued that 
$\theta$ is in general an independent critical exponent, not related by 
scaling laws to other critical exponents, although the relation 
(\ref{scaling}) is exact for $n=\infty$, to first order in $\epsilon = 
4-d$, and for the $d=1$ Ising model (for which the dynamics are Markovian), 
and seems to be numerically quite accurate for the $d=2$ Ising model. 
The corresponding exponent for the global order parameter following a quench 
into the ordered phase is also of interest, and is currently under 
investigation by numerical simulations. 

AB and SC's research was supported by EPSRC (UK), SM's by NSF grant 
no.\ DMR-92-24290. AB thanks SISSA (Italy), and Amos Maritan in 
particular, for hospitality during the final stages of this work, and 
Alan McKane for discussions.

\end{multicols}
\end{document}